\def\beg{\begin{equation}}
\def\eeq{\end{equation}}
\documentstyle[12pt]{article}
\begin{document}
\baselineskip18pt
\begin{center}
{\Large{\bf A NEW PHASE IN THE BILAYERS OF SEMICONDUCTORS IN QUANTUM HALL EFFECT}}
\vskip0.35cm
{\bf KESHAV N. SHRIVASTAVA}\\
{\it School of Physics, University of Hyderabad, \\
Hyderabad  500046, India}
\end{center}
\vskip0.5cm
We find that a bilayer of semiconductors emits a new Goldstone
quasiparticle when Landau levels in the layers are half filled.
The emission of the new quasiparticle is associated with the
divergence in the energy of the system characteristic of a phase
transition.\\

keshav@mailaps.org          Fax. +91-40-3010145\\
\baselineskip24pt
\noindent{\bf 1.~~Introduction.}\\

Recently, Spielman et al$^{1,2}$ have found a new type of
Goldstone boson in quantum Hall effect in bilayers. The
Goldstone boson is accompanied with a phase transition or broken
symmetry. In the vacuum state if there is a broken symmetry,
then a massless particle is emitted. Spielman et al have
discovered such a quasiparticle. However, it remains to be
proved that there is a phase transition associated with the new
quasiparticle. Usually the Hall resistivity is quantized in
units of $h/ie^2$ where $i$ is an integer. It has been reported
that $i$ need not be an integer. In particular, it can be equal
to 1/3 or some other fraction. We have found$^3$ that the
experimentally observed fractions can be accurately predicted by
a special spin-orbit interaction based on a pseudoscalar which
describes a $\vec{v}.\vec{s}$ type interaction, where $\vec{v}$
is the velocity of the electron and $\vec{s}$ the spin. Thus
there is a new spin-dependent force on the electron. This
spin-orbit interaction is of the order of $v/c$ whereas the
usual spin-orbit interaction in atoms is of the order of
$v^2/c^2$. It has been found experimentally that the masses of
some of the quasiparticles are equal. This equality of masses
has been understood by using the particle-hole symmetry.$^4$
Further details of this calculation have been described in
detail$^5$. The effective charge and the high Landau levels are
well understood by our calculation$^6$.

In the present paper, we show that at half filled Landau level,
the energy of the system diverges so that there is a phase
transition. This phase transition is of second order and hence
it is accompanied by a new type of Goldstone boson. In the
present case, boson is emitted when $\sqrt{2}$ times the
magnetic length is an integer multiple of the distance between
the two layers of the semiconductors.
\vskip0.5cm
\noindent{\bf 2.~~Theory.}

We show that when a flux is moving between two walls at a
distance $d$ apart, a sharp peak in the conductance is predicted
when magnetic length, $t$, is an integer multiple of $d$,
otherwise a broad peak occurs. Using an earlier work, we show
that there is a phase transition when both layers are having
half-filled Landau levels. This phase is accompanied by a
Goldstone type boson. When the magnetic field is applied in the
$z$ direction and the electric field along $y$ direction, Hall
voltage develops along the $x$ direction. The Hall resistivity
is linearly proportional to the applied magnetic field and can
be used to determine the carrier concentration, $n$ as,
\beg
\rho_H = {B\over nec}
\eeq
where $B$ is the magnetic induction and $e$ is the charge of the
electron with $c$ as the velocity of light. In 1980 von Klitzing
et al$^5$ discovered that there is a plateau at integer
multiples of $h/ie^2$ ($i$=integer), so that 
\beg
\rho_{xy} = {h\over ie^2}.
\eeq
Equating (1) and (2) we find that
\beg
 {B\over nec} = {h\over ie^2}\,\,.
\eeq
We define the concentration as the number of electrons per unit
area, $n=n_o/A$ so that the above relation becomes
\beg
BA = {n_o\over i} (hc/e)
\eeq
which shows that flux is quantized with unit charge $e$. The
magnetic field may be measured in terms of a distance such that
the area $A=t^2$ above, may be taken as the square of the
magnetic length, $t$ so that (4) may be written as,
\beg
Bt^2 = {1\over i} (hc/e)
\eeq
where we have used $n_o=1$ as the number of electrons per unit
area. For $i=1$,
\beg
Bt^2 = hc/e
\eeq
and
\beg
t = \left( {hc\over eB}\right)^{1/2}
\eeq
which is called the magnetic length. In the case of a two layer
system with distance $d$ between layers, the following theorems
are predicted. For
\beg
t = id
\eeq
sharp line in the conductance versus voltage is predicted where
magnetic length is perfectly matched with the distance between
semicodncutor layers. When $t<<d$ or $t>>d$, the sharp line
disappears and a broad line is then predicted. Thus for a sharp
line in the characteristic conductance,
\beg
Bi^2d^2 = {hc\over e}
\eeq
otherwise a broad line is predicted. Since the integer $i$ has
come from the area, the square of this integer is generated. The
fraction of conductivity, $\sigma=dI/dV$, in the sharp line is
\beg
\sigma_o = \sigma_t {\sin^2(\pi d/t)\over(\pi d/t)^2}\,\,,
\eeq
where $\sigma_o$ is contained in the sharp line and $\sigma_t$
is the total integrated value of the conductivity. This spectrum
of $dI/dV$ as a function of $V$ is equivalent to the motion of
vortex waves with $t$ as the wave length. Since the condition
(9) gives a Goldstone mode, there must be a phase transition
associated with zero of an order parameter or the divergence of
a physical quantity, such as susceptibility of a ferromagnet. We
will show that the energy of the system diverges at half filled
Landau level. Since the fractional charges predicted by us$^3$
are exactly the same as those experimentally observed by
Eisenstein and St\"ormer$^7$ and others,$^{8-11}$ we make use of
this model to understand the divergence. All of the aspects of
our model$^3$ are in agreement with the experimental data$^{7-11}$.

We have reported that a special type of spin-orbit interaction
predicts the fractions of charges correctly. This interaction is
of first order in $v/c$ whereas the ordinary spin-orbit coupling
of an electron in an atom is of second order, varying as
$v^2/c^2$. The spin dependence in the interaction is introduced
by the product $\vec{s}.\hat{n}$ where $\hat{n}$ is a unit
vector in the direction of the radius vector $\vec{r}$. From the
vectors $\vec{s}.\vec{n}$ and $\vec{v}$, a true scalar of the
form $\hat{n}\times\vec{v}.\vec{s}$ may be constructed. The new
spin-orbit coupling operator is of the form, 
\beg
V_{sl} = - \phi(r) \hat{n} \times\vec{v}.\vec{s}
\eeq
where $\phi(r)$ is a function of $\vec{r}$ because
$\vec{l}=\vec{r}\times \vec{p}$. This interaction may be written
as a spin-orbit interaction,
\beg
V_{sl} = -f(r)\vec{l}.\vec{s}
\eeq
where $f(r)=\hbar\phi(r)/rm$. It splits the levels with orbital
angular momentum $l$ into two levels, $j=l\pm1/2$. Since,
\begin{eqnarray}
-\vec{l}.\vec{s} &=& -{1\over2} l \qquad\mbox{for}\qquad j=l+{1\over2}\nonumber\\
	          &=& +{1\over2} (l+1) \qquad\mbox{for}\qquad j=l-{1\over2}
\end{eqnarray}
the energy difference between the two states is 
\begin{eqnarray}
\Delta E &=& E_{(l-{1\over2})} - E_{(l+{1\over2})}\nonumber\\
&=& f(r) (l+ {1\over2})
\end{eqnarray}
As $l\to\infty$, one of the levels with $j=l+{1\over2}$ in (13)
goes to $-\infty$ and the other with $j=l-{1\over2}$ goes to
$+\infty$ and the energy different in (14) always goes to
$\infty$. Therefore, there is a phase transition as
$l\to\infty$. Therefore the Goldstone mode is accompanied by a
phase transition. Next, we show that at this phase transition
the Landau level is half filled. We obtain the effective
fractional charge from the charge in the Bohr magneton or from
the cyclotron frequency. Usually $l=0$, for the conduction
electrons but we have to allow the finite values of $l$ to
obtain the required result. We consider the spin as well as the
orbital motion so that,
\beg
g_j\vec{j}=g_s\vec{s} + g_l\vec{l} = {1\over2}(g_l+g_s)\vec{j} +
{1\over2}(g_l-g_s)(\vec{l}-\vec{s})\,\,\,.
\eeq
We consider the bound electrons which have finite $l$. Upon
substituting $s={1\over2}$, the above expression gives,
\beg
g_j = g_l \pm {g_s-g_l\over2l+1}
\eeq
which for $j=l\pm1/2$ and $g_s=2$, $g_l=1$ gives,
\beg
g_\pm = 1\pm {1\over2l+1}\,\,\,.
\eeq
The cyclotron frequency is defined in terms of the magnetic
field as,
\beg
\omega = {eB\over mc}\,\,\,.
\eeq
Corresponding to this frequency, the voltage along $y$ direction
is
\beg
\hbar\omega=eV_y\,\,\,\,.
\eeq
From the above two relations
\beg
{e^2B\over2\pi mc} = {e^2\over h} V_y 
\eeq
which describes the current in the $x$ direction so that the
resistivity is
\beg
\rho_{xy} = {h\over e^2}
\eeq
which is the same as (2) for $i=1$. We include (17) in (18)
so that the current becomes,
\beg
I_x = {1\over2} g{e^2B\over2\pi mc} = {1\over2} {ge^2V_y\over h}
\,\,\,\,.
\eeq
For $l=0$, $g=2$ and,
\beg
I_x = {e^2\over h} V_y
\eeq
so that we define the effective charge as,
\beg
\nu_\pm = {1\over2} g_\pm\,\,\,.
\eeq
For $l=0, {1\over2}g_+=1$ and ${1\over2}g_-=0$, for $l=1,
{1\over2} g_+ = {2\over3}$ and ${1\over2}g_-={1\over3}$ which
determines the fractional charge of a quasiparticle. From (17)
and (24) we show below the predicted effective charge,
\vskip0.25cm
\begin{center}
\begin{tabular}{lcccccccc}
\hline
$l$                  & 0 & 1   & 2   & 3   & 4   & 5    & 6    & $\infty$\\
\\
$\nu_-={1\over2}g_-$ & 0 & 1/3 & 2/5 & 3/7 & 4/9 & 5/11 & 6/13 & 1/2\\
\\
$\nu_+={1\over2}g_+$ & 1 & 2/3 & 3/5 & 4/7 & 5/9 & 6/11 & 7/13 & 1/2\\
\hline
\end{tabular}
\end{center}
\vskip0.25cm
\noindent which occurs in two series such that for a given value
of $l$ the sum of the two values is always equal to unity, 
\beg
\nu_+ + \nu_- = 1
\eeq
because of the Kramers conjugate pairs, one has spin + and the
other -. At $l=\infty$, $g_\pm=1$ for both the series and hence
$\nu_\pm={1\over2}$ which corresponds to half filled Landau
level. Thus we find that at $l=\infty$, the Landau level is half
filled, and as we have seen below eq.(14), there is a divergence
and hence phase transition. The modified values of the magnetic
length is
\beg
t^* = \left({hc\over\nu eB}\right)^{1/2} = id
\eeq
for the Goldstone mode which for $\nu={1\over2}$ is at,
\beg
\left({2hc\over eB}\right)^{1/2} = id\,\,\,.
\eeq
Thus the divergence in energy at $l=\infty$, $\nu={1\over2}$
which describes a phase transition, is accompanied by a Goldstone
boson. We have obtained three different features, (a) a
Goldstone boson, (b) divergence in energy and (c) the correct
fractional charge by using the $\vec{v}.\vec{s}$ type
interaction given by (11) alone. We can write the velocity in
the form of a current as $\vec{J}=n_ee\vec{v}$ where $n_e$ is
the concentration of electrons so that the relevent interaction becomes,
\beg
{\cal H}^\prime = -{\phi(r)\over n_ee} \hat{n}\times\vec{J}.\vec{s}
\eeq
which describes all of the essential features of the 
quantum Hall effect correctly.
\vskip0.5cm
\noindent{\bf 3.~~Comparison with the experimental data}.

Usually the Goldstone boson in solids may be a soft phonon
associated with a lattice distortion. The zero-phonon lines are
common in solids. The M\"ossbauer lines are also zero-phonon
lines. However, in the present problem of semiconductor
heterostructures, there are no phonons and the problem is
electromagnetic at low temperatures and hence these are new type
of Goldstone bosons of intensity
\beg
I \propto \exp(-d^2/t^2)\,\,\,.
\eeq
The predicted values of the effective fractional charge given
above eq.(25) are exactly the same as those experimentally
observed by Eisenstein and St\"ormer.$^7$ Not only the numerical
values but also the grouping into two groups predicted here is
in agreement with that observed. Similarly, the predicted values
are in agreement with the experimentally observed values in many
other measurements. When $\nu$ is an observable fractional
charge, $n\nu$ ($n=$ integer) also becomes observable. Since
$l=\infty$ limit gives $\nu=1/2$, the values $n/2$ become
observable. This prediction is also in accord with the
experimental data of Yeh et al.$^{11}$ The divergence in energy
at $\nu=1/2$ shows that there is a phase transition at these
values in the semiconductor bilayer to a new state of matter.
The symmetry breaking at this point is accompanied with a
Goldstone boson. This prediction of emission of a boson is in
agreement with the experimental observation of a Goldstone boson
at zero voltage$^{1-2}$.

We have predicted that there is a Goldstone boson at $\nu=1/2$
when the condition (27) is satisfied in the bilayer
semiconductor. This mode is associated with a divergence in
energy at $l\to\infty$ as determined from (14). The predicted
fractional charges are in agreement with the experimental data
in several works$^{7-11}$. The predicted Goldstone boson at the
phase transition is also in agreement with that found
experimentally. A Josephson type flux flow has recently been
considered$^{12,13}$ in the semiconductor bilayers which is not
in contradiction with our work because we can convert the flux to current. Several advanced theories of the
quantum Hall effect have been reviewed from which we find that
our theory is consistent with those of others$^{14}$. The fractions 
given by Fig. 18 of St\"ormer$^{15}$ are exactly the same as those
predicted by us$^{3}$. There is no doubt in our interpretation of the
Goldstone boson observed by measuring the conductance in the data
of Spielman et al$^{1}$. The conductance is analogous to the
Doppler brodened line calculated by Dicke$^{16}$. However,
in the case of Dicke's line there are ordinary distances whereas
in the present case there is the magnetic length. The interaction
(11) involves the vector product of the velocity with the unit vector
in the direction of the radius vector which generates the rotations
or "magnetic rotons" which are often observed in the experiments. Thus
the Goldstone boson, divergence in energy at $l=1/2$ required
for a singularity in the energy, effective fractional charges and 
rotations, predicted by our calculations are all in good agreement
with the data.\\ 
\noindent{\bf 4.~ Bose condensation or phase transition.}

The equation (24) gives the effective fractional charge. The
series with + sign gives $\nu_+=1/2$ as $l\to\infty$. Similarly,
$\nu_-=1/2$ occurs also at $l=\infty$. We call these states
$A^{(+)}$ and $B^{(-)}$, respectively. The energy of the state
$A^{(+)}$ from (13) is $-(1/2)l$ and that of $B^{(-)}$ is
$+{1\over2}(l+1)$. When the quasiparticles from these states are
added to form a mixed state, the energy is
$[-{1\over2}l+{1\over2}(l+1)]f(r)={1\over2}f(r)$, i.e., we
consider $\epsilon_1$ and $\epsilon_2$ as the single particle
energies for the two states and form a two particle state with
energy, $\epsilon_1+\epsilon_2$. The Bose distribution for the
mixed state is,

\begin{equation}
n = {1\over e^{(\epsilon_1+\epsilon_2-\mu)/k_BT}-1}
\end{equation}
where $\mu$ is the chemical potential of the mixed state arising
from the interactions between the $A^{(+)}$ and $B^{(-)}$ states.
So far we have not introduced the wave vector space and hence
$\epsilon_1$ and $\epsilon_2$ are not associated with any wave
vectors but $\epsilon_1$ has spin + and $\epsilon_2$ has spin -.
Therefore $\epsilon_1+\epsilon_2$ actually is similar to a Cooper
pair with infinite value of $l$. The two particle energy is, 
$$
\epsilon_1+\epsilon_2 = {\hbar^2k^2_1\over2m} + {\hbar^2k^2_2\over2m}
+ {1\over2} f(r) + \hbar\omega_c(n_1+{1\over2})+\hbar\omega_c(n_2+{1\over2})
$$
where the last two terms are the result of Landau levels when the
Landau number in state $\epsilon_1$ need not be equal to that in
$\epsilon_2$ state. The minimum energy is obtained at $k_1=k_2=0$
and $n_1=n_2=0$,
\begin{equation}
2\epsilon_o=(\epsilon_1+\epsilon_2)_{min} =
{1\over2}f(r)+\hbar\omega_c\,\,\,.
\end{equation}
The number of particles in the ground state is
\begin{equation}
n_o={1\over e^{(2\epsilon_o-\mu)/k_BT}-1}\,\,\,.
\end{equation}
If these particles are confined in a 2-dimensional plane, the
average number of particles is given by,
\begin{equation}
N=n_o+{A\over(2\pi)^2} \int
{kdk\over e^{(\epsilon_{1k}+(\epsilon_{2k}-\mu)/k_BT} -1}\,\,\,.
\end{equation}
We devide this expression by the area $A$ to obtain the number of
particles per unit area as,
\begin{equation}
{N\over A} = {n_o\over A} + {1\over(2\pi)^2} \int {kdk\over e^{(\epsilon_{1k}+\epsilon_{2k}-\mu)/k_BT}-1}
\end{equation}
which diverges for $\epsilon_{1k}+\epsilon_{2k}=\mu$. This
divergence in the areal density is associated with the Bose
condensation. We write $(\epsilon_{1k}+\epsilon_{2k}-\mu)/k_BT=x$
and $(\hbar^2k^2_1+\hbar^2k_2^2)/2m\simeq ak^2$ so that
\begin{equation}
k_BTx+\mu=ak^2\,\,\,.
\end{equation}
Differentiating this expression,
\begin{equation}
kdk = {k_BT\over 2a} dx
\end{equation}
which substituted in the above gives the number of quasiparticles
per unit area as,
\begin{equation}
{N\over A} = {n_o\over A} + {(k_BT)/2a\over(2\pi)^2} \int {dx\over e^x-1} \,\,\,.
\end{equation}
This means that there is a divergence in $N/A$ when $x=0$ and
away from the divergence, $N/A$ depends linearly on temperature.
The particles in the states $A^{(+)}$ and $B^{(-)}$ may Bose
condense in the two particle states and the resulting liquid
exhibits superflow and hence can tunnel between layers just like
a Josephson tunneling. As we have pointed out this Bose condensed
state is made from pairs of quasiparticles with one electron spin
up and the other spin down but not $s$ wave because of large $l$. Therefore, the pairs
can tunnel through the layers of GaAs/AlGaAs as in Josephson
effect. However, in the Josephson effect in superconductors, the
pairs are usually in the $s$ wave, $l=0$ and in some materials a
$d$-wave has been found with $l=2$. In the present problem
$\nu={1over2}$ is achieved with $l\to\infty$. Hence, the present
state has $l\to\infty$ which is never the case in
superconductors. The $l\to\infty$ state with spin singlet state
is a boson but in superconductors, the Cooper pairs are made from
fermions. The eq.(24) gives the effective fractional charge. The
series with + sign gives $\nu_+=1/2$ at $l\to\infty$. Similarly
$\nu_-=1/2$ occurs alos at $l=\infty$. We call these states as
$A^+$ and $B^-$, respectively. The energy of the state $A^+$ from
(13) is $-(1/2)l$ and that of $B^-$ is ${1\over2}(l+1)$. The
Bose-Einstein distribution for the $A^+$ state is,
\begin{equation}
n={1\over e^{(\epsilon_A-\mu)/k_BT}-1}
\end{equation}
where the single particle energy in the state $A^+$ is,
\begin{equation}
\epsilon_A = {\hbar^2k^2\over2m} - {1\over2}l f(r)+\hbar\omega_c (n+{1\over2}).
\end{equation}
At $\nu={1\over2}$, $\l\to\infty$ and hence $A^+$ state energy
diverges so that there is a phase transition when single-particle
$A^+$ and $B^-$ states are formed. This phase transition is
associated with a Goldstone mode and not with Bose condensation.
The paired state with $l\to\infty$, with one particle spin up and
the other down is a Bose condensed state. Therefore, the two
particle state, one spin up and the other down, with $l=\infty$
for both, gives a Bose condensed state which is not a phase
transition. The single particle state at $l=\infty$ does not Bose
condense but gives the phase transition and hence the Goldstone
mode. The two particle state gives the Bose condensed state. Thus
two results emerge. Firstly, the two particle state gives Bose
condensation and in this case the Bose condensate itself has to
be its own Goldstone mode. Secondly, the single particle
$\nu=1/2$ does not Bose condense but describes a phase transition
and hence a Goldestone mode emerges.\\
     We have compared our theory with a number of data and found that
in all cases the theory is in accord with the experimental data$^{17}$.
We find that there is a new change in the magnetic moment of the electron$^{18}$ in the quantum Hall effect. We find that the rate of sweep of the magnetic field plays an important role particularly in NMR$^{19}$ and even if NMR is not being taken. The theory of the composite fermions (CFs)in found to be internally inconsistent$^{20}$ so there are no CFs in the real material. The polarization of the electrons as a function of temperature exhibits ordinary two-level type behavior and shows that fluxex are {\it not} attached to the electrons$^{21}$. 
\vskip0.5cm
\noindent{\bf 5.~~Conclusions}

The interaction with dot product of velocity with spin, given by
(11), exerts a new force on the electron. The energies arising
from this interaction show that there is a divergence and hence
a phase transition so that there is a new type of Goldstone
boson. We predict that the quasiparticles exhibit effective
fractional charge. The predicted Goldstone boson and the
fractional charges are in agreement with the experimental observations.
There is a Bose condensation of a two component fluid at $\nu = 1/2$.\\
\baselineskip18pt
\noindent{\bf References}
\begin{enumerate}
\item I.B. Spielman, J.P. Eisenstein, L.N. Pfeiffer and K.W.
	West, Phys. Rev. Lett. {\bf84}, 5808 (2000).
\item I.B. Spielman, J.P. Eisenstein, L.N. Pfeiffer and K.W.
	West, Cond-mat/0012094. 
\item K.N. Shrivastava, Phys. Lett. A{\bf113}, 435 (1986);
	{\bf115}, 459(E) (1986).
\item K.N. Shrivastava, Mod. Phys. Lett. {\bf13}, 1087 (1999).
\item K.N. Shrivastava, Superconductivity: Elementary Topics,
	World Scientific, Singapore (2000).
\item K.N. Shrivastava, Mod. Phys. Lett. {\bf14}, 1009(2001).
\item J.P. Eisenstein and H.L. St\"ormer, Science {\bf248}, 1510 (1990).
\item R.R. Du, H.L. St\"ormer, D.C. Tsui, L.N. Pfeiffer and K.W.
	West, Phys. Rev. Lett. {\bf70}, 2944 (1993).
\item W. Pan, J.S. Xia, V. Shavarts, D.E. Adams, H.L. St\"ormer,
	D.C. Tsui, L.N. Pfeiffer, K.W. Baldwin and K.W. West,
	Phys. Rev. Lett. {\bf83}, 3530 (1999).
\item J.P. Eisenstein, M.P. Lilly, K.B. Cooper, L.N. Pfeiffer
	and K.W. West, Physica E{\bf6}, 29 (2000).
\item A.S. Yeh, H.L. St\"ormer, D.C. Tsui, L.N. Pfeiffer, K.W.
	Baldwin and K.W. West, Phys. Rev. Lett. {\bf62}, 592
	(1999). 
\item M.M. Fogler and F. Wilczek, Phys. Rev. Lett. {\bf86}, 1833
	(2001).
\item M.M. Fogler and F. Wilczek, cond-mat/0007403.
\item M. Stone, ed., Quantum Hall effect, World Scientific,
	Singapore 1992.
\item H. L. St\"ormer, Rev. Mod. Phys. {\bf71}, 875 (1999).
\item R. H. Dicke, Phys. Rev.  {\bf89}, 472 (1953).
\item K. N. Shrivastava, CERN SCAN-0103007.
\item K. N. Shrivastava, cond-mat/0104004.
\item K. N. Shrivastava, cond-mat/0104577.
\item K. N. Shrivastava, cond-mat/0105559.
\item K. N. Shrivastava, cond-mat/0106160.
\end{enumerate}

\end{document}